\documentclass[12pt]{iopart}
\expandafter\let\csname equation*\endcsname\relax
\expandafter\let\csname endequation*\endcsname\relax

\usepackage{epsfig,xfrac,subfigure,booktabs,xfrac,amsmath,graphicx}

\newcommand\sr{\mathrm{sr}}
\newcommand\ini{\mathrm{ini}}

\newcommand\sca{\mathrm{S}}
\newcommand\ten{\mathrm{T}}

\newcommand\ret{\mathrm{ret}}
\newcommand\D{\mathrm{d}}
\newcommand\ele{\mathrm{l}}
\newcommand\ere{\mathrm{r}}

\newcommand\phai{\mathrm{pi}}
\newcommand\ua{\mathrm{ua}}
\newcommand\im{\mathrm{i}}

\begin{document}

\title{Semiclassical analysis of the tensor power spectrum in the Starobinsky inflationary model}

\author{Truman Tapia}

\address{$^1$Yachay Tech University, School of Physical Sciences and Nanotechnology, Hda. San Jos\'e s/n y Proyecto Yachay, 100119, Urcuqu\'i, Ecuador}

\author{Clara Rojas}
\address{$^2$Yachay Tech University, School of Physical Sciences and Nanotechnology, Hda. San Jos\'e s/n y Proyecto Yachay, 100119, Urcuqu\'i, Ecuador}
\ead{crojas@yachaytech.edu.ec}

\begin{abstract}
In this work we calculate the tensor power spectrum and the tensor-to-scalar ratio $r$  within the frame of the Starobinsky inflationary model using the improved uniform approximation method and the third-order phase-integral method. We compare our results with those obtained with numerical integration and the slow-roll approximation to second order. We have obtained  consistent values of $r$ using the different approximations, and $r$ is inside the interval reported by observations.
	
\end{abstract}
 
\maketitle

\section{Introduction}

Inflation was introduced by Alan Guth in the eighties to solve the fine-tuning problems \cite{guth:1981}. Inflation is an early stage of the Universe in which an accelerated expansion occurred \cite{liddle:2000}.  Plenty of inflationary models have been proposed, which makes the discrimination of these models an active field of research. Scalar and tensor power spectra, and their parameters, are the main predictions made by inflation models. These predictions are compared with observations to test the models. The scalar power spectrum has been widely used for this purpose, but coming projects will measure parameters related with tensor perturbations \cite{matsumura2014mission,collaboration2011core,kogut2016primordial}. Then, accurate calculations of the tensor power spectrum are meaningful presently. 

The Starobinsky inflationary model was introduced in the eighties \cite{barrow:1988a,barrow:1988b,starobinsky:1980}  and has caused interest in recent years \cite{linde:2014,diValentino:2017,paliathanasis:2017,adam:2019,granada:2019,samart:2019,chowdhury:2019,renzi:2020,truman:2020}.
This model is well supported by the recent observation of the Planck 2018 results \cite{akrami:2018}, and it is used is plenty theoretical studies in cosmology currently. The classical predictions of this model to lowest order in the slow-roll approximation are, for the spectral index $n_s - 1\approx -2/N$, and for the tensor-to-scalar ratio $r\approx 12/N^2$.

Slow-roll approximation is one of the most used methods to study the perturbation power spectra in the context of inflation. Another method is to solve numerically mode by mode the equations of the dynamics of perturbations (scalar and tensor). In recent years, semi-classical methods have appeared in the literature as an alternative way to study the equation of perturbations and calculate the power spectrum. Our goal in this work is to calculate the tensor power spectrum and the tensor-to-scalar ratio using semiclassical methods such as, the improved uniform approximation method  \cite{habib:2002,casadio:2005,casadio:2006} and the third-order phase-integral approximation \cite{rojas:2007b,rojas:2007c,rojas:2009,rojas:2012,truman:2020}. The results will be compared with the numerical approximation and the slow-roll approximation to second order. These semiclassical methods are faster than the calculations made with a numerical code and produce results comparable with the exact result.

The article is structured as follows: In Section 2 we show how we address the background dynamics of the Universe in the context of the Starobinsky model. In Section 3 we describe the equations of perturbations and the power spectra. In Section 4 we discuss the different approximation methods employed in this work. Section 5 shows the results that we obtained. Finally, in Section 6 we make some conclusions.
 
\section{Dynamics in the Starobinsky model}

\begin{figure}[th!]
\centering
\includegraphics[scale=0.35]{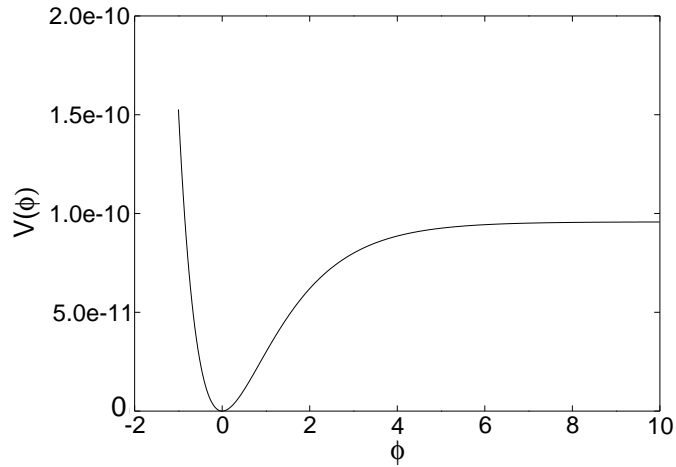}
\caption{The Starobinsky potential.}
\label{potential}
\end{figure}

The Starobinsky potential is given by \cite{martin:2014,martin:2019}

\begin{equation}
\label{V}
V(\phi)=M^4 \left(1-e^{-\sqrt{\sfrac{2}{3}}\,\phi}\right)^2,
\end{equation}
where $\phi$ is the scalar field driven inflation, and $M=3.13 \times 10^{-3}$ \cite{mishra:2018}.
In Fig. \ref{potential} we show the Starobinsky potential, at first sight  the potential is sufficiently flat to produce inflation.

The Friedmann and the continuity equations dictate the dynamics of the Universe govern by a single scalar field $\phi$, these equations are:

\begin{equation}
\label{back1}
    H^2=\dfrac{1}{3}\left[V(\phi)+\dfrac{1}{2}\dot{\phi}^2\right],
\end{equation}

\begin{equation}
\label{back2}
    \ddot{\phi}+3H\dot{\phi}=-V,_\phi,
\end{equation}

\noindent where $H=\dot{a}/a$ is the Hubble parameter. Given that analytical solutions for these equations are not available, we solve them numerically. The initial conditions employed are $a_0=1$, $\phi_0=5.77$, and $\phi'_0=-1.04\times 10^{-7}$, which guarantees enough inflation to solve the fine-tuning problems, and the correct amplitude of the scalar power spectrum. These numerical solutions will be used to solve the equations of perturbations shown in the next section. For the semiclassical methods we use the following fits of the numerical solutions of $a$ and $\phi$:

\begin{eqnarray}
\nonumber
a_\textnormal{fit}&=&2.71828^{(-3.53339 + 5.65 \times 10^{-6} t)} (111.184 - 7.53333 \times 10^{-6} t)^{\sfrac{3}{4}},\\\\
\phi_\textnormal{fit}&=&1.2309 \ln\left(108.59- 7.29215 \times 10^{-6}  t\right).
\end{eqnarray}

\section{Equations of perturbations}

The scalar perturbations are described by the function $u=a\Phi/\phi'$, where $\Phi$ is a gauge-invariant variable corresponding to the Newtonian potential. The equations of motion of the perturbation $u_k$ in Fourier space are

\begin{equation}\label{dotdotuk}
u_k''+\left(k^2-\dfrac{z_{S}''}{z_{S}}\right)u_k=0,
\end{equation}
where $z_{S}=a\phi'/\mathcal{H}$, $\mathcal{H}=a'/a$, and the prime indicates derivative with respect to the conformal time $\eta$.

For tensor perturbations one introduces the function $v_k=ah$, where $h$ represents the amplitude of the gravitational
wave. Tensor perturbations obey  a second order differential equation analogous to Eq. (\ref{dotdotuk}):

\begin{equation}\label{dotdotvk}
v_k''+\left(k^2-\dfrac{a''}{a}\right)v_k=0.
\end{equation}
Considering the limits  $k^2\gg|z_{S}''/z_{S}|$ (short wavelength) and $k^2\ll|z_{S}''/z_{S}|$  (long wavelength), we have that  the  solutions to
Eq. (\ref{dotdotuk}) exhibit the following asymptotic behavior:

\begin{equation}
\label{boundary_0}
u_k\rightarrow \dfrac{e^{-ik\eta}}{\sqrt{2k}}
\quad \left(k^2\gg|z_{S}''/z_{S}|, -k\eta\rightarrow \infty \right),
\end{equation}

\begin{equation}
\label{boundary_i} u_k\rightarrow A_k z  \quad \left(k^2\ll|z_{S}''/z_{S}|,
-k\eta\rightarrow 0\right).
\end{equation}

\noindent Equation \eqref{boundary_0} is used as the initial condition for the perturbations. The same asymptotic conditions hold for tensor
perturbations.

Once the solutions for $u_k$ and $v_k$ are known, the power spectra for scalar and tensor perturbations are given by the expressions

\begin{eqnarray}
\label{PS}
P_\sca(k)&=& \lim_{-k\eta\rightarrow 0} \dfrac{k^3}{2 \pi^2}\left|\dfrac{u_k(\eta)}{z_{S}(\eta)} \right|^2,\\
\label{PT}
P_\ten(k)&=& \lim_{-k\eta\rightarrow 0} \dfrac{k^3}{2 \pi^2}\left|\dfrac{v_k(\eta)}{a(\eta)} \right|^2.
\end{eqnarray}

\noindent Also, the tensor-to-scalar ratio $r$ is defined as \cite{habib:2005b}

\begin{equation}
\label{R}
r=8\dfrac{P_T(k)}{P_S(k)}.
\end{equation}

\noindent The spectral indices for scalar and tensor perturbations are defined by:

\begin{eqnarray}
\label{nS}
n_\sca(k)&=& 1+\dfrac{\D\ln P_\sca(k)}{\D\ln k},\\
\label{nT}
n_\ten(k)&=&  \dfrac{\D\ln P_\ten(k)}{\D\ln k}.
\end{eqnarray}

Since the scale factor $a$ and the field $\phi$ exhibit a simpler form in the physical time $t$ than in the conformal time $\eta$, we proceed to write the equation for the tensor perturbations in the variable $t$. The relation between $t$ and $\eta$ is given via the equation $\D t= a\D \eta$. In this case, the equation for tensor perturbations can be written as

\begin{equation}
\label{dotvk}
\ddot{v_k}+\dfrac{\dot{a}}{a}\dot{v_k}+\dfrac{1}{a^2}\left[k^2-\left(\dot{a}^2+a\ddot{a}\right) \right]v_k=0.
\end{equation}

\noindent This equation is solved to find the numerical solution discussed in more detail in the next section. Also, in order to apply the semiclassical methods we use this equation with the change of variable $v_k(t)=\frac{V_k(t)}{\sqrt{a}}$, obtaining that $V_{k}$ satisfies the differential equation:

\begin{equation}
\label{ddotVk}
\ddot{V}_k+R_\ten(k,t)V_k=0,
\end{equation}
where

\begin{equation}
\label{RT}
R_\ten(k,t)=\dfrac{1}{a^2}\left[k^2-\left(\dot{a}+a\ddot{a}\right)\right]+\dfrac{1}{4a^2}\left(a^2-2a\ddot{a}\right),
\end{equation}

\medskip
\noindent
$R_\ten(k,t)$ is  calculated numerically and $V(k)$ satisfies the asymptotic conditions

\begin{eqnarray}
\label{bordeVk}
V_k&\rightarrow&\sqrt{\dfrac{a(t)}{2k}}\exp{\left[-ik\eta(t)\right]}, \quad k\,t\rightarrow 0,\\
\label{ceroVk}
V_k&\rightarrow&A_k \sqrt{a(t)}\, a(t),\quad  k\,t\rightarrow \infty.
\end{eqnarray}

\begin{figure}[th!]
\centering
\includegraphics[scale=0.35]{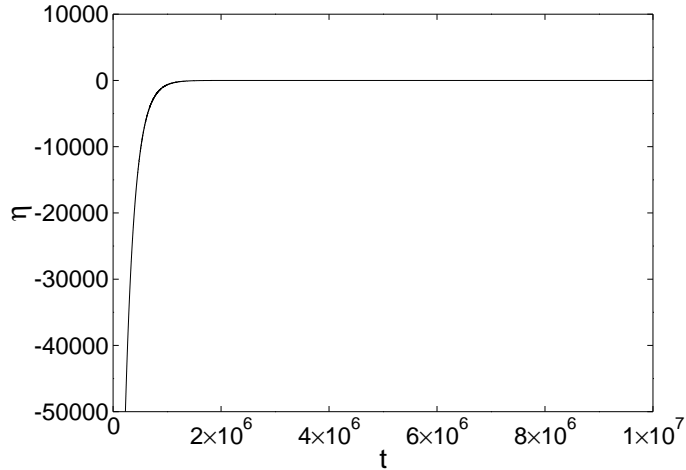}
\caption{Conformal time $\eta$ as a function of the physical time $t$ for the Starobinsky inflationary model.}
\label{eta}
\end{figure}

In order to apply  the asymptotic condition \eqref{boundary_0} and \eqref{bordeVk}, we  use the relation between $\eta$ and $t$ given by:

\begin{equation}
\D\eta=\int_{t_\ini}^t \dfrac{\D t}{a(t)},
\end{equation}
where  $t_\ini=10^7$, then $\eta$ is zero at the end of the inflationary epoch. Figure \ref{eta} shows $\eta$ as a function of $t$, and we can observe that:
\begin{eqnarray}
\textnormal{when} \quad -k\,\eta\rightarrow 0          &\Rightarrow&  k\,t \rightarrow \infty,\\
\textnormal{when} \quad -k\,\eta\rightarrow \infty  &\Rightarrow& k\,t \rightarrow 0.
\end{eqnarray}

\medskip

Eq. \eqref{dotvk}, does not  possess an exact analytical solution. In order to solve this differential equation we make use of the methods: numerical integration, slow-roll approximation up-to second-order, the improved uniform approximation, and the third-order phase integral approximation. Then, we compare the results of these different methods.

\section{Solutions of the perturbation equation}

\subsection{Numerical Integration}
The equation for tensor perturbations \eqref{dotvk} is integrated numerically. To set this equation, we use the results coming from numerical integration of the equations \eqref{back1} and \eqref{back2}. The perturbation $v_k$ is a complex function, then two differential equations are solved, one for the real part and the other for the imaginary part. The strategy to solve these two equations is basically the same. 

The integration is done in two parts, one in the limit when $k^2\gg \dot{a}^2+a\ddot{a}$, and the other when the full equation \eqref{dotvk} is considered. The first part corresponds with the time when perturbations are inside the horizon, then $v_k$ exhibits an oscillatory behavior. In this limit we solve the reduced equation

\begin{equation}
\label{dotvk_k2}
\ddot{v_k}+\dfrac{\dot{a}}{a}\dot{v_k}+\dfrac{k^2}{a^2} v_k=0,
\end{equation}
from 300 to 100 oscillations before the horizon crossing, using as initial condition equation \eqref{boundary_0}. Then, we use the final stage of this solution as initial condition, to solve equation \eqref{dotvk} from 100 oscillations before horizon crossing to roughly three times the horizon crossing time when the perturbation is frozen. The numerical integration was done using the Software Mathematica version 12.1, the code is available\footnote[1]{link}. 

\subsection{Slow-roll approximation}

The tensor power spectrum in the slow-roll approximation up-to second-order is given by the expression \cite{stewart:2001}

\begin{eqnarray}
\label{PT_sr}
\nonumber
P_\ten^{\sr}(k)&\simeq\Big[1+(2b-2)\epsilon_1+\left(2b^2-2b-3+\dfrac{\pi^2}{2}\right)\epsilon_1^2\\
&+\left(-b^2+2b-2+\dfrac{\pi^2}{12}\right)\epsilon_2\Big]\left.\left(\dfrac{H}{2\pi}\right)^2\right|_{k=aH},
\end{eqnarray}
where  $b=0.729637$ is the Euler constant, and

\begin{eqnarray}
\epsilon_1&=&-\dfrac{\dot{H}}{H^2},\\
\epsilon_2&=&\dfrac{1}{H}\dfrac{\D\epsilon_1}{\D t}.
\end{eqnarray}

\noindent Also, the spectral index and the tensor-to-scalar ratio \cite{casadio:2006} in the slow-roll approximation are given by:

\begin{eqnarray}
\label{nT_sr}
\nonumber
n_\ten^{\sr}(k)&\simeq&-2\epsilon_1-2\epsilon_1^2+(2b-2)\epsilon_2,\\
r & \simeq & 16\, \epsilon_1\left(1+C \epsilon_2\right),
\end{eqnarray}
where $C=-0.7296$.
\noindent
All the expressions in the slow-roll approximation must be evaluated at the time $t_*$, which is the horizon crossing time when $k=aH$. By using different values of $k$ in the range $0.0001\, \textnormal{Mpc}^{-1}  \leq k \leq 10\, \textnormal{Mpc}^{-1}$, we can obtain the $k$ dependence of the tensor power spectrum.

\subsection{Uniform approximation}

We want to obtain an approximate solution to the differential equation (\ref{ddotVk})  in the range where  $Q_\ten^2(k,t)$ have a simple root  at $t_\ret=\upsilon_\ten$, so that $Q_{\ten}^2(k,t)>0$ for  $0<t<t_\ret$ and $Q_{\ten}^2(k,t)<0$ for  $t>t_\ret$ as depicted in Fig. \ref{QTa}. Using the uniform approximation method \cite{berry:1972,habib:2002,rojas:2007b,rojas:2007c,rojas:2009}, we obtain that for  $0<t<t_\ret$ 

\begin{eqnarray}
\label{Ukzero}
\nonumber
V_k(k,t)&=&\left[\dfrac{\rho_\ele(k,t)}{Q_\ten^2(k,t)} \right]^{1/4} \left\{C_1
A_i[-\rho_\ele(k,t)]+C_2 B_i[-\rho_\ele(k,t)] \right\},\\
\dfrac{2}{3}\left[\rho_\ele(k,t)\right]^{3/2}&=&\int_{t}^{t_\ret} \left[Q_{\ten}^2(k,t)\right]^{1/2}\D t,
\end{eqnarray}
\medskip
where  $C_1$ and  $C_2$ are two constants to be determined with the help of the boundary conditions (\ref{bordeVk}). 
For $t> t_\ret$

\begin{eqnarray}
\label{Vkinfinity}
\nonumber
V_k(k,t)&=&\left[\dfrac{-\rho_\ere(k,t)}{Q_\ten^2(k,t)} \right]^{1/4} \left\{C_1 A_i[\rho_\ere(k,t)]+ C_2 B_i[\rho_\ere(k,t)] \right\},\\
\dfrac{2}{3}\left[\rho_\ere(k,t)\right]^{3/2}&=&\int_{t_\ret}^{t} \left[-Q_{\ten}^2(k,t)\right]^{1/2}\D t,
\end{eqnarray}

For the computation of the power spectrum we need to take the limit $k\,t\rightarrow \infty$ of the solution  (\ref{Vkinfinity}). In this limit we have

\begin{eqnarray}
\label{limit_vk}
\nonumber
v_k^\ua(t)&\rightarrow&  \dfrac{C}{\sqrt{2\,a(t)}}\left[-Q_\ten^2(k,t)\right]^{-1/2}\\
\nonumber
&\times& \left\{ \dfrac{1}{2}\exp\left(-\int_{\upsilon_\ten}^{t}\left[-Q_\ten^2(k,t)\right]^{1/2} \D t\right)+\im\,\exp\left(\int_{\upsilon_\ten}^{t}\left[-Q_\ten^2(k,t)\right]^{1/2} \D t\right)\right\},\\
\end{eqnarray}
where $C$ is a phase factor.  Using  the growing part  of the solutions   (\ref{limit_vk}),  one can compute the tensor  power spectrum using the uniform approximation method,

\begin{eqnarray}
P_\ten(k)&=&\lim_{-k t\rightarrow \infty} \dfrac{k^3}{2\pi^2} \left|\dfrac{v_k^\ua(t)}{a(t)}\right|^2.
\end{eqnarray} 
 
We are going to use the second-order improved uniform approximation for the power spectrum \cite{rojas:2009,habib:2005b}

\begin{equation}
 \tilde{P}_\ten(k)=P_\ten(k)\left[\Gamma^*(\bar{\nu}_\ten)\right]^2,
\end{equation}
where $\bar{\nu}_{\sca,\ten}$ is the turning point for the tensor power spectrum and 

\begin{equation}
\Gamma^*(\nu)\equiv 1+\dfrac{1}{12\nu}+\dfrac{1}{288\nu^2}-\dfrac{139}{51840\nu^3}+\cdots.
\end{equation}

\subsection{Phase-integral approximation}
\begin{figure}[htbp]
\begin{center}
\subfigure[]{
\label{QTa}
\includegraphics[scale=0.3]{QT.eps}}
\subfigure[]{
\label{QTb}
\includegraphics[scale=0.3]{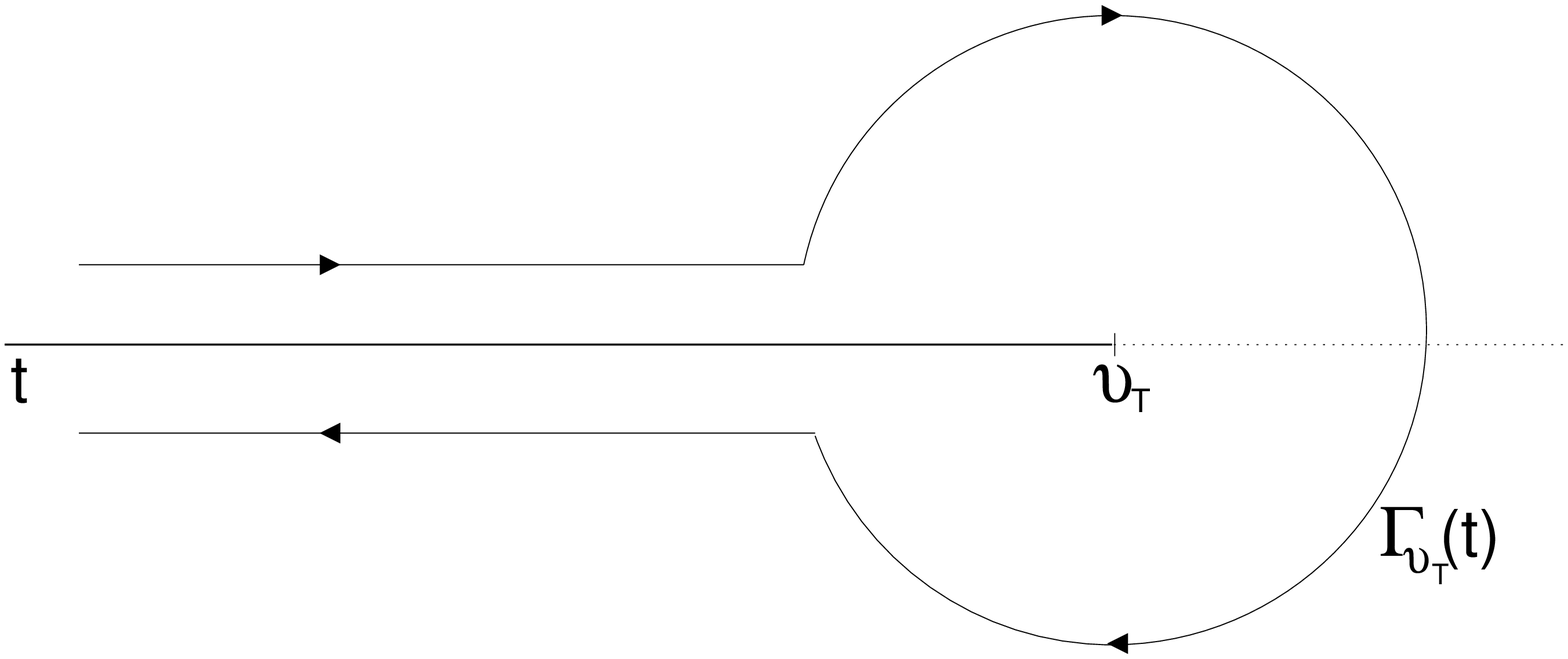}}
\subfigure[]{
\label{QTc}
\includegraphics[scale=0.3]{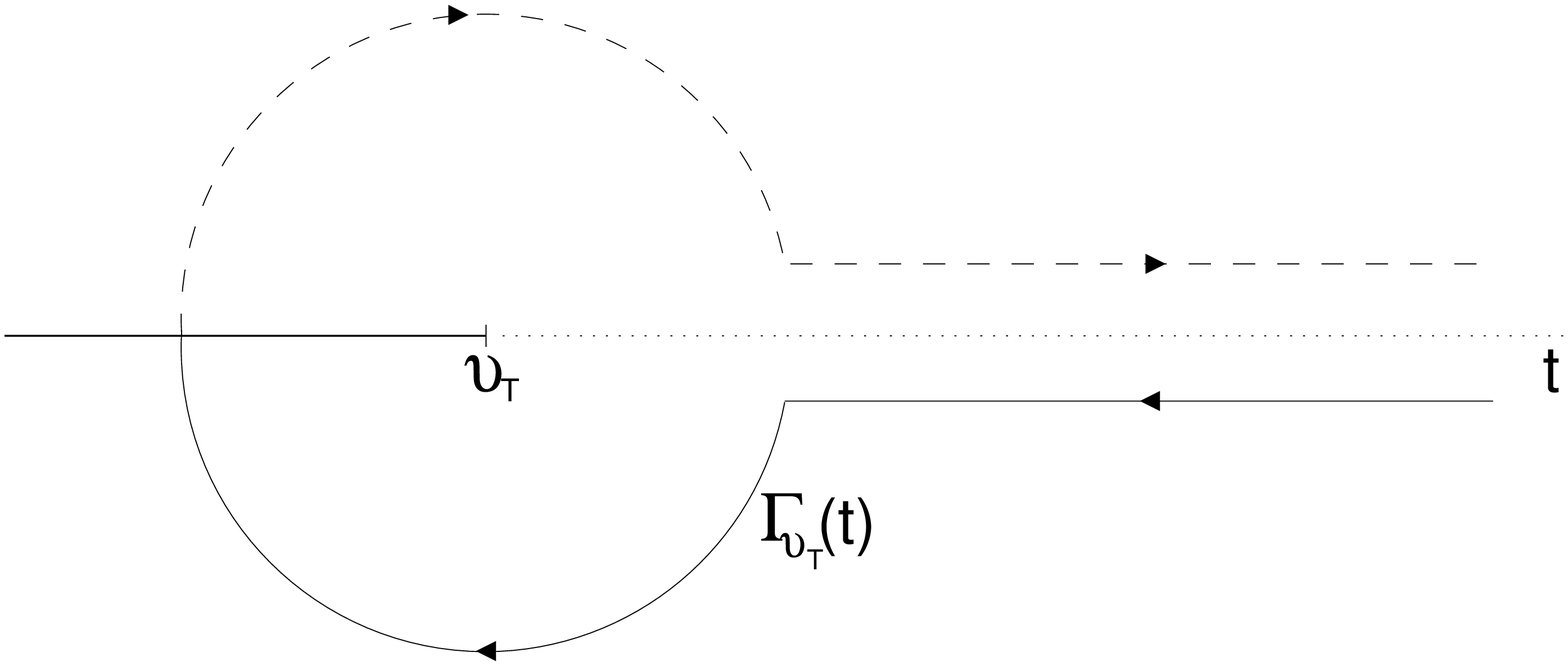}}
\caption{\small{(a) Behaviour of  $Q_\ten^2(k,t)$.
(b) Contour of integration $\Gamma_{\nu_\ten}(t)$ for $0<t<\nu_\ten$. 
(c) Contour of integration  $\Gamma_{\nu_\ten}(t)$ for $t>\nu_\ten$. 
The dashed lined indicates the part of the path on the second Riemann sheet.}}
	\end{center}
\end{figure}

In order to solve Eq. (\ref{ddotVk})  with the help of the phase-integral approximation \cite{froman:1996}, we choose the following base functions $Q_\ten$ for the tensor  perturbations

\begin{eqnarray}
\label{Q}
Q_\ten^2(k,t)&=&R_\ten(k,t),
\end{eqnarray}
where  $R_\ten(k,t)$ is given by  Eq. (\ref{RT}). Using this selection, the phase-integral approximation is  valid as  $k t\rightarrow \infty$, and in this limit we should impose the condition (\ref{ceroVk}), where the validity condition   $\mu \ll 1$ holds.  The bases functions  $Q_\ten(k,t)$  possess turning points, for each mode $k$ this turning points represent the horizon. There are two ranges  where to define the solution. To the left of the turning point  $0<t<t_\ret$ we have the classically permitted region  $Q_{\ten}^2(k,t)>0$ and to the right of the turning point $t>t_\ret$ corresponding to the classically forbidden region $Q_{\ten}^2(k,t)<0$, such as it is shown in Figs \ref{QTa}.

The mode  $k$ equations for the tensor perturbations (\ref{ddotVk})  in the phase-integral approximation has two solutions: 

\noindent For $0<t <t_\ret$

\begin{eqnarray}
\label{ukleft}
\nonumber
v^\phai_k(t)&=& \dfrac{d_1}{\sqrt{a(t)}}\left|q_\ten^{-1/2}(k,t)\right| \cos{\left[\left|\omega_\ten(k,t)\right|-\dfrac{\pi}{4}\right]} \\
&+& \dfrac{d_2}{\sqrt{a(t)}}\left|q_\ten^{-1/2}(k,t)\right| \cos{\left[\left|\omega_\ten(k,t)\right|+\dfrac{\pi}{4}\right]}.
\end{eqnarray}

\noindent For $t>t_\ret$

\begin{eqnarray}
\label{ukright}
\nonumber
v^\phai_k(t)&=&\dfrac{d_1}{2\sqrt{a(t)}}\left|q_\ten^{-1/2}(k,t)\right|\exp\left[-\left|\omega_\ten(k,t)\right|\right]\\
&+& \dfrac{d_2}{\sqrt{a(t)}} \left|q_\ten^{-1/2}(k,t)\right| \exp\left[\left|\omega_\ten(k,t)\right|\right].
\end{eqnarray}

\noindent Using the phase-integral approximation up to third order ($2N+1=3\rightarrow N=1$), we have that $q_{\ten}(k,t)$  can be expanded in the form

\begin{eqnarray}
\nonumber
q_\ten(k,t)&=&\sum_{n=0}^1 Y_{2n_\ten}(k,t) Q_\ten(k,t),\\
\label{q1}
&=&\left[Y_{0_\ten}(k,t)+Y_{2_\ten}(k,t)\right] Q_\ten(k,t).
\end{eqnarray}

In order to compute $q_\ten(k,t)$, we  compute $Y_{2_\ten}(k,t)$, and the required function  $\varepsilon_{0_\ten}(k,t)$.  The expression  (\ref{q1}) gives  a third-order approximation for $q_\ten(k,t)$.  In order to compute $\omega_\ten(k,t)$  we make a contour integration following the path indicated in Fig. \ref{QTb}-(c).

\begin{eqnarray}
\omega_\ten(k,t)&=&\omega_{0_\ten}(k,t)+ \omega_{2_\ten}(k,t),\\
&=&\int_{\upsilon_\ten}^{t}Q_\ten(k,t)\D t+\dfrac{1}{2}\int_{\Gamma_{\upsilon_\ten}}Y_{2_\ten}(k,t)Q_\ten(k,t)\D t,\\
&=&\int_{\upsilon_\ten}^{t}Q_\ten(k,t)\D t+\dfrac{1}{2}\int_{\Gamma_{\upsilon_\ten}}f_{2_\ten}(k,t)\D t,
\end{eqnarray}
\medskip
where

\begin{eqnarray}
f_{2n_\ten}(k,t)&=&Y_{2_\ten}(k,t)Q_\ten(k,t).
\end{eqnarray}
The function $f_{2_\ten}(k,t)$  have the following functional dependence:

\begin{eqnarray}
\label{A}
f_{2_\ten}(k,t)&=&A(k,t)(t-\upsilon_\ten)^{-5/2},
\end{eqnarray}
where $A(k,t)$ is regular at $\upsilon_\ten$. With the help of  the function  (\ref{A}) we compute the integrals for $\omega_{2n}$  up to $N=1$ using the contour indicated in  Fig. \ref{QTb}-(c). The expressions for  $\omega_{2n}$ permit one to obtain the third-order phase integral approximation of the solution to the equation for tensor perturbations (\ref{ddotVk}).  The constants $d_1$ and $d_2$ are obtained using the limit  $k\,t\rightarrow 0$ of the solutions on the left side of the turning point (\ref{ukleft}), and  are given by the expressions

\begin{eqnarray}
d_1&=&-\im\,d_2,\\
d_2&=&\dfrac{\e^{-\im\dfrac{\pi}{4}}}{\sqrt{2}}\e^{-\im\left[k\,\eta(0)+\left|\omega_{0_\ten}(k,0)\right|\right]}.
\end{eqnarray}

\medskip
In order to compute the tensor power spectrum, we need to calculate the limit  $k\,t\rightarrow \infty$ of the growing part of the solutions on the right side of the turning point  given by   Eq. (\ref{ukright})

\begin{eqnarray}
P_\ten(k)&=&\lim_{-k t\rightarrow \infty} \dfrac{k^3}{2\pi^2} \left|\dfrac{v_k^\phai(t)}{a(t)}\right|^2.
\end{eqnarray}

\section{Results}

\begin{figure}[th!]
\centering
\includegraphics[scale=0.35]{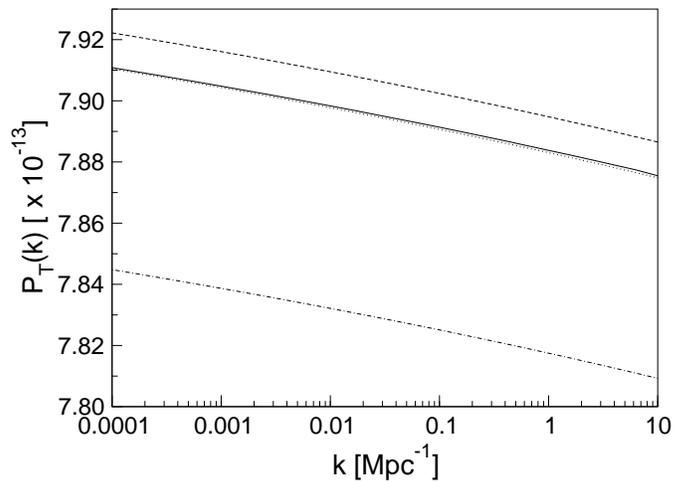}
\caption{$P_\ten(k)$  for the Starobinsky inflationary model.  Solid line: numerical result; dashed line: third-order phase-integral approximation; dot-dashed line: second-order improved uniform approximation, dotted line: second-order slow-roll approximation.}
\label{PT_graph}
\end{figure}

\begin{figure}[th!]
\centering
\includegraphics[scale=0.35]{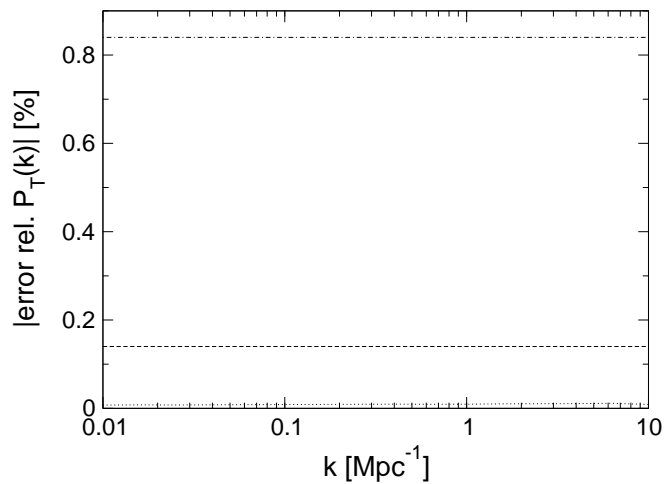}
\caption{Relative error with respect to the numerical result of $P_\ten(k)$   for the Starobinsky inflationary model.  Dashed line: third-order phase-integral approximation; dot-dashed line: uniform approximation; dotted line: second-order slow-roll approximation.}
\label{error_PT}
\end{figure}

With the initial conditions and parameters shown in Section 2 we get that the scalar power spectrum $P_\sca=2.09577 \times 10^{-9}$ for the mode $k=0.05 Mpc^{-1}$, which agrees with measurements reported by Planck 2018 results\cite{akrami:2018}. To rescale the momentum $k$ in physical units we have followed the work of Habib \textit{et al.} \cite{habib:2005b}. Figure \ref{PT_graph} shows $P_\ten (k)$ using each method described in the last section. We can observe that the semiclassical methods work very well and they are of easier implementation than the numerical method. Fig. \ref{error_PT} shows the relative error of each  approximation method with respect to the numerical result. The second-order slow roll approximation deviates in $0.008\%$, the improved uniform approximation deviates from the numerical result in $0.84\%$, whereas that the phase-integral approximation up to third-order deviates $0.14\%$.

The value of the tensor spectral index for each approximation method is given in Table 1. Also, the values of the tensor-to-scalar ratio are presented in Table 2. We have found that of the approximation methods used in this work the better value of $r=0.00272633$  is obtained with  the third-order phase-integral method, which is inside the upper limit reported by Planck $2018$ results \cite{akrami:2018,kehagias:2014}. 

\begin{table}[th!]
\begin{tabular}{ccc}
\toprule
Method                                     & $n_\ten(k)$           & rel. err (\%)\\ 
\midrule
Numerical                                  & $\;\;-0.000393769$     & $\;\;$              \\
Second-order slow roll                     & $\;\;-0.000394629$    & $\;\;0.2184$   \\
Improved Uniform approximation   & $\;\; -0.000396719$     & $\;\;0.7492$ \\
Phase integral method up to third-order    & $\;\; -0.000396287$     & $\;\;0.6395$   \\
\bottomrule
\end{tabular}
\caption{Value of  $n_\ten$  obtained  with different approximation methods for the Starobinsky inflationary model at the pivot scale $k=0.05$ Mpc$^{-1}$.}
\label{graph_nT}
\end{table}

\begin{table}[th!]
\begin{tabular}{ccc}
\toprule
Method                                     & $r(k)$           & rel. err (\%)\\ 
\midrule
Numerical                                  & $\;\;0.00272608$     & $\;\;$              \\
Second-order slow roll                     & $\;\;0.00282388$    & $\;\;3.5879$   \\
Improved Uniform approximation    & $\;\; 0.00273398$     & $\;\;0.2897$ \\
Phase integral method up to third-order    & $\;\; 0.00272633$     & $\;\;0.009171$   \\
\bottomrule
\end{tabular}
\caption{Value of  tensor-to-scalar ratio $r$  obtained  with different approximation methods for the Starobinsky inflationary model at the pivot scale $k=0.002\, \textnormal{Mpc}^{-1}$.}
\label{graph_r}
\end{table}

\section{Conclusions}

We have shown that the semiclassical methods are useful tools to calculate  the power spectrum of tensor perturbations and its subsequent parameters $r$ and $n_T$. With the phase-integral method we have obtained a value of $r$ with a relative error of $0.009171 \%$, it is worth to mention that the values obtained for tensor-to-scalar ratio are in the interval reported by Planck $2018$ results \cite{akrami:2018}.



\bibliographystyle{unsrt}

\begin{thebibliography}{10}

\bibitem{guth:1981}
A.~H. Guth.
\newblock {Inflationary universe: A possible solution to the horizon and
  flatness problems}.
\newblock {\em Phys. Rev. D}, 23:347, 1981.

\bibitem{liddle:2000}
{A. R. Liddle and D. H. Lyth}.
\newblock {\em {Cosmological inflation and large-scale structure}}.
\newblock Cambridge University Press, 2000.

\bibitem{matsumura2014mission}
{T. Matsumura\textit{et al.}}
\newblock {Mission design of LiteBIRD}.
\newblock {\em Journal of Low Temperature Physics}, 176:733, 2014.

\bibitem{collaboration2011core}
{COrE Collaboration: C. Armitage-Caplan \textit{et al.} }.
\newblock Core (cosmic origins explorer) a white paper.
\newblock {\em arXiv:1102.2181}, 2011.

\bibitem{kogut2016primordial}
{A. Kogut, J. Chluba, D. Fixsen, S. Meyer, and D. Spergel}.
\newblock {The primordial inflation explorer (PIXIE)}.
\newblock In {\em Space Telescopes and Instrumentation 2016: Optical, Infrared,
  and Millimeter Wave}, volume 9904, page 99040W. International Society for
  Optics and Photonics, 2016.

\bibitem{barrow:1988a}
{John D. Barrow}.
\newblock {The Premature Recollapse Problem in Closed Inflationary Universes}.
\newblock {\em Nucl. Phys. B}, 296:697, 1988.

\bibitem{barrow:1988b}
{John D. Barrow and S. Cotsakis}.
\newblock {Inflation and the Conformal Structure of Higher Order Gravity
  Theories}.
\newblock {\em Phys. Lett. B}, 214:515, 1988.

\bibitem{starobinsky:1980}
A.~A. Starobinsky.
\newblock {A new type of isotropic cosmological models without singularity}.
\newblock {\em Phys. Lett. B}, 91:99, 1980.

\bibitem{linde:2014}
A.~Linde.
\newblock {Inflationary Cosmology after Planck 2013}.
\newblock {\em arXiv:1402.0526}, 2014.

\bibitem{diValentino:2017}
{E. Di Valentino and L. Mersini-Houghton}.
\newblock {Testing predictions of the quantum landscape multiverse 1: the
  Starobinsky inflationary potential}.
\newblock {\em JCAP}, 2, 2017.

\bibitem{paliathanasis:2017}
A.~Paliathanasis.
\newblock {Analytic solution of the Starobinsky model for inflation}.
\newblock {\em Eur. Phys. J C}, 77:438, 2017.

\bibitem{adam:2019}
{C. Adam and D. Varela}.
\newblock {The superpotential method in cosmological inflation}.
\newblock {\em arXiv:1901}, 2019.

\bibitem{granada:2019}
{L. N. Granada and D. F. Jimenez}.
\newblock {Slow-roll inflation with exponential potential in scalar-tensor
  models}.
\newblock {\em Eur. Phys. J. C}, 79:772, 2019.

\bibitem{samart:2019}
{D. Samart and P. Channuie}.
\newblock {Unification of inflation and dark matter in the Higgs-Starobinsky
  model}.
\newblock {\em Eur. Phys. J. C}, 79:347, 2019.

\bibitem{chowdhury:2019}
{D. Chowdhury, J. Martin, C. Ringeval, and V. Vennin}.
\newblock {Inflation after Planck: Judgment Day}.
\newblock {\em arXiv:1902.03951}, 2019.

\bibitem{renzi:2020}
{F. Renzi, M. Shokri, and A. Melchiorri}.
\newblock {What is the amplitude of the gravitational waves background expected
  in the Starobinsky model?}
\newblock {\em Phys. Dark Univ.}, 27:100450, 2020.

\bibitem{truman:2020}
{T. Tapia, M. Z. Mughal, and C. Rojas}.
\newblock {Semiclassical analysis of the Starobinsky inflationary model}.
\newblock {\em Phys. Dark Univ.}, 30:100650, 2020.

\bibitem{akrami:2018}
{Y. Akrami \textit{et al.}}
\newblock {Planck 2018 results. X. Constraints on inflation}.
\newblock {\em arXiv:1807.06211}, 2018.

\bibitem{habib:2002}
{S. Habib and A. Heinen and K. Heitmann and G. Jungman and C. Molina-Par\'is}.
\newblock {The Inflationary Perturbation Spectrum}.
\newblock {\em Phys. Rev. Lett.}, 89:281301, 2002.

\bibitem{casadio:2005}
{R. Casadio, F. Finelli, M. Luzzi, and G. Venturi}.
\newblock {Improved WKB analysis of cosmological perturbations}.
\newblock {\em Phys. Rev. D}, 71:043517, 2005.

\bibitem{casadio:2006}
{R. Casadio, F. Finelli, A. Kamenshchik, M. Luzzi, and G. Venturi}.
\newblock {The method of comparison equations for cosmological perturbations}.
\newblock {\em JCAP}, 04:011, 2006.

\bibitem{rojas:2007b}
{Clara Rojas and V\'ictor M. Villalba}.
\newblock {Computation of inflationary cosmological perturbations in the
  power-law inflatioary model using the phase-integral method}.
\newblock {\em Phys. Rev. D}, 75:063518, 2007.

\bibitem{rojas:2007c}
{V\'ictor M. Villalba and Clara Rojas}.
\newblock {Applications of the phase integral method ins ome inflationary
  scenarios}.
\newblock {\em J. Phys. Conf. Ser.}, 66:012034, 2007.

\bibitem{rojas:2009}
{Clara Rojas and V\'ictor M. Villalba}.
\newblock {Computation of inflationary cosmological perturbations in chaotic
  inflationary scenarios using the phase-integral method}.
\newblock {\em Phys. Rev. D}, 79:103502, 2009.

\bibitem{rojas:2012}
{Clara Rojas and V\'ictor M. Villalba}.
\newblock {Computation of the power spectrum in chaotic
  $\frac{1}{4}\lambda\phi^4$ inflation}.
\newblock {\em JCAP}, 003:1, 2012.

\bibitem{martin:2014}
{J. Martin, C. Ringeval, and V. Vennin}.
\newblock {Encyclopaedia Inflationaris}.
\newblock {\em Phys. Dark Univ.}, 5-6:75--235, 2014.

\bibitem{martin:2019}
J.~Martin.
\newblock {Cosmic Inflation: Trick or Treat?}
\newblock {\em arXiv:1902.05286}, 2019.

\bibitem{mishra:2018}
{S. S. Mishra, V. Sahni, and A. V. Toporensky}.
\newblock {Initial conditions for inflation in an FRW universe}.
\newblock {\em Phys. Rev. D}, 98:083538, 2018.

\bibitem{habib:2005b}
{S. Habib and A. Heinen and K. Heitmann and G. Jungman}.
\newblock {Inflationary Perturbations and Precision Cosmology}.
\newblock {\em Phys. Rev. D}, 71:043518, 2005.

\bibitem{stewart:2001}
{E. D Stewart and J. Gong}.
\newblock {The density perturbation power spectrum to second-order corrections
  in the slow-roll expansion}.
\newblock {\em Phys. Lett. B}, 510:1, 2001.

\bibitem{berry:1972}
{M. Berry and K. E. Mount}.
\newblock {Semiclassical Approximations in Wave Mechanics}.
\newblock {\em Rep. Prog. Phys.}, 35:315, 1972.

\bibitem{froman:1996}
{N. Fr\"oman and P. O. F\"oman}.
\newblock {\em {Phase-Integral Method. Allowing Nearlying Transition Point}},
  volume~40.
\newblock Springer Tracts in Natural Philosophy, 1996.

\bibitem{kehagias:2014}
{A. Kehagias, A. M. Dizgah, and A. Riotto}.
\newblock {Remarks on the Starobinsky model of inflation and its descendants}.
\newblock {\em Phys. Rev. D}, 89:043527, 2014.

\end{thebibliography}

\end{document}